\documentclass[prl,twocolumn,english,showpacs,nofootinbib]{revtex4-1}
\usepackage{amssymb}
\usepackage{color}
\usepackage{graphicx}
\usepackage{bbold}
\usepackage{bm}
\usepackage{amsmath}
\usepackage{amsfonts}
\usepackage{amssymb}
\usepackage{braket}
\usepackage{units}
\usepackage{csquotes}
\usepackage{upgreek}
\usepackage{xcolor}
\usepackage{fancyhdr}
\usepackage{float}
\usepackage{cancel}
\begin{document}
	
%	\pagestyle{fancy}
%	\chead{draft \today}
%	\rhead{\thepage}
%	\renewcommand{\headrulewidth}{0.0pt} 

\title{Pair correlations in the attractive Hubbard model}

\author{C. F. Chan, M. Gall,  N. Wurz, and M. K\"ohl}
\affiliation{Physikalisches Institut, University of Bonn, Wegelerstra{\ss}e 8, 53115 Bonn, Germany}

\begin{abstract}
{The mechanism of fermionic pairing is the key to understanding various phenomena such as high-temperature superconductivity and the pseudogap phase in cuprate materials. We study the pair correlations in the attractive Hubbard model using ultracold fermions in a two-dimensional optical lattice. By combining the fluctuation-dissipation theorem and the compressibility equation of state, we extract the interacting pair correlation functions and deduce a characteristic length scale of pairs as a function of interaction and density filling. At sufficiently low filling and weak on-site interaction, we observe that the pair correlations extend over a few lattice sites even at temperatures above the superfluid transition temperature.} 
\end{abstract}
\maketitle

The nature of fermionic pairing plays a decisive role in many-body quantum states such as superconductors \cite{Lee2006}. For attractive interactions and in the weak-coupling regime, a Bardeen-Cooper-Schrieffer (BCS) description gives rise to large, spatially-overlapping Cooper pairs below the critical temperature $T_c$ \cite{Singer1996}. Upon increasing attractive interaction, the BCS ground state crosses over to a Bose-Einstein condensate (BEC) of tightly-bound dimers \cite{Zwerger2011bcs}. Above the transition temperature $T_c$,  the crossover of a normal state could be characterized by the formation of preformed pairs, which are linked to the emergence of a pairing pseudogap \cite{Randeria1992AttHub, Ding1996, EsslingerReview2010}. In a lattice configuration, the pairing phenomenon can be described in simple terms using the Hubbard model with an attractive $s$-wave interaction. Understanding the formation of preformed pairs offers a key insight into more complicated pairing mechanisms such as the $d$-wave pairing in high-temperature superconductors. 

Quantum mechanically the pairing behavior between two particles is fully encoded in the second-order pair correlation function $g^{(2)}$. In position space, $g^{(2)}(\boldsymbol{r})$ gives the joint probability of finding a pair of particles spaced by distance $\boldsymbol{r}$. The pair correlation is of particular importance as it is related to various thermodynamical observables including compressibility, pressure and internal energy. In solid materials,  $g^{(2)}(\boldsymbol{r})$ is inferred from the Fourier transform of the density structure factor $S(\boldsymbol{q})$, which is experimentally accessible by crystallographic methods such as X-ray diffraction and neutron scattering \cite{Garman2015Review,bee1988quasielastic}. For ultracold atomic gases in optical lattices that emulate the Hubbard model, the occurrence of spatial density and spin correlations with repulsive interactions were detected via high-resolution microscopy or scattering experiments \cite{Cheuk2016b, Drewes2016, Mazurenko2016,brown2017spin, Hart2015, Drewes2017}. More recently, the s-wave pairing pseudogap in the attractive Hubbard model was accessed via photoemission spectroscopy in a quantum gas microscope \cite{brown2019angle}. 

In this work, we investigate the formation of pair correlations in the two-dimensional attractive Hubbard model. In particular, we measure the thermodynamic correlation function and make use of the fluctuation-dissipation theorem to obtain an estimate of the pair correlation length. We observe the formation of pairs with increasing attractive interactions above the superfluid critical temperature. The pair correlation length is of great importance as it can distinguish between a pseudogap pairing phase and a Fermi liquid phase above the critical temperature. While past experiments have determined the pair size in a trapped Fermi superfluid across the BCS-BEC crossover  \cite{schunck2008determination}, it has not been measured for lattice gases. 

\begin{figure}[t!]
\includegraphics[width=0.475\textwidth]{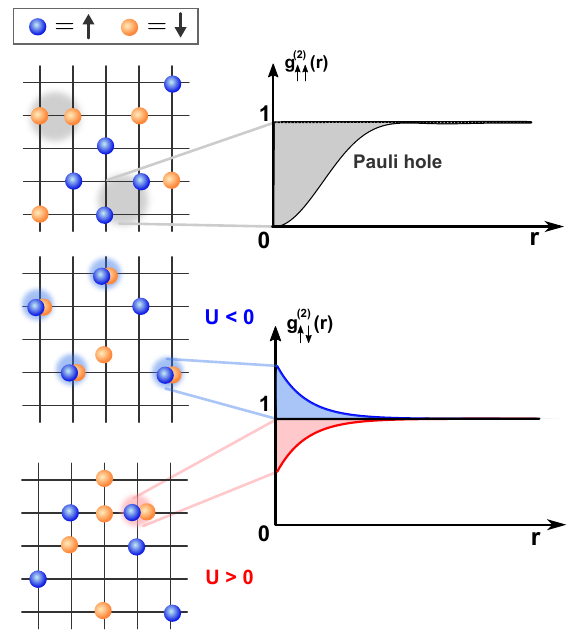}
\caption{Schematic diagram for pairing in a two-dimensional square lattice. The equal-spin correlation function $g_{\uparrow \uparrow}^{(2)}(\textbf{r})$ exhibits fermionic anti-bunching as a correlation hole. The unequal-spin correlation function $g_{\uparrow \downarrow}^{(2)}(\boldsymbol{r})$ depends strongly on the on-site interaction strength $U$, where attractive (repulsive) interaction raises (lowers) pair correlation at short distance from the classical value of one.} 
\label{Fig1}
\end{figure}

Due to the Pauli principle, density correlations exist even for ideal fermions. The equal-spin pair correlation function $g_{\uparrow\uparrow}^{(2)} (\boldsymbol{r}) $ exhibits a Pauli correlation hole at short distance, as shown in Fig.~\ref{Fig1}. This gives rise to an effective repulsion that solely originates from the anti-symmetric nature of the fermionic state. Particles with opposite spin, however, are uncorrelated in an ideal Fermi gas and thus $g_{\uparrow\downarrow}^{(2)} (\boldsymbol{r}) = 1$. For repulsively interacting fermions, the pair correlation between particles with opposite spin is suppressed with respect to the classical value of one \cite{Drewes2016}, which goes hand in hand with the emergence of the Mott insulator. For attractive interactions, the enhancement of the pair correlation signals the formation of pairs.

% For interacting fermions, the pair correlation between particles with opposite spin is suppressed by a repulsive interaction \cite{Drewes2016} and enhanced by an attractive one. While the former goes hand in hand with the emergence of the Mott insulating phase, the latter signals the formation of local pairs. 

%Moreover, the band filling $n$ imposes an additional density dependence on the $g^{(2)}$ function. For increasing filling, the pair correlation functions at all non-zero distance converges to 1. This is best illustrated in the limiting case of band insulator ($n=2$), where the probability of finding a particle pair is independent of distance,  \textit{i.e.} $g_{\sigma\sigma}^{(2)} (\boldsymbol{r}) =g_{\sigma\sigma'}^{(2)} (\boldsymbol{r}) = 1$, except for identical fermions at $\boldsymbol{r} = 0$. 

The static density structure factor $S(\boldsymbol{q})$ for constant filling $n$, which incorporates density fluctuations at all length scales, is linked to the Fourier transform of the pair correlation function, and is given by
\begin{equation}
S(\boldsymbol{q}) =  1 +  n \displaystyle{\int} \left[g^{(2)}(\boldsymbol{r}) - 1\right] e^{-i \boldsymbol{q}  \cdot \boldsymbol{r} } d^2r .
\label{densitySF}
\end{equation}
Here, $g^{(2)}(\boldsymbol{r})$ denotes the full pair correlation function $ g^{(2)}(\boldsymbol{r})= \frac{1}{4} \sum_{\sigma = \uparrow, \downarrow}  \left(g^{(2)}_{\sigma \sigma}(\boldsymbol{r}) + g^{(2)}_{\sigma \sigma'}(\boldsymbol{r}) \right)$ \cite{schwabl1999advanced}. 
Although theoretically obtaining the full $g^{(2)}$ from Eq.~(\ref{densitySF}) requires access to the density structure factor across all momenta, the limiting case at $\boldsymbol{q}=0$ still encapsulates both the Pauli blocking and interacting contributions of $g^{(2)}(\boldsymbol{r})$. In addition, the fluctuation-dissipation theorem connects the density structure factor at zero momentum to the corresponding thermodynamical susceptibility, in this case, the isothermal compressibility $\kappa = \left( \partial n / \partial \mu \right)|_T$, \textit{i.e.} $S(\boldsymbol{q}=0) = \kappa T/n$ \cite{Ho2009,Drewes2016}. Combining this with Eq. (\ref{densitySF}) yields the relation
\begin{equation}
\displaystyle{\int}\left[ g_{\uparrow\uparrow}^{(2)} (\boldsymbol{r}) + g_{\uparrow\downarrow}^{(2)} (\boldsymbol{r}) - 2 \right] d^2r   = 2 \left( \frac{\kappa T}{n^2} -\frac{1}{n} \right).
\label{g2Full}
\end{equation}
Eq.~(\ref{g2Full}) provides us with two important insights. First, the isothermal compressibility entails the competition of Pauli repulsion and on-site attraction, since an external compression induces pressure between particles with equal and unequal spins within the trap. Second, the experimental determination of the compressibility opens access to the pair correlation function. While previous quantum gas experiments have been focusing on measurements of compressibility in relation to transport coefficients such as conductivity \cite{brown2019bad}, its link to pairing has yet to be explored. 

% Experimental realizations
Our experiment realizes the two-dimensional Hubbard model with attractive interaction on a square lattice, which reads
\begin{equation}
H = -t \sum_{\langle i,j \rangle, \sigma}  c^\dagger_{i,\sigma} c_{j,\sigma} + U \sum_{i} n_{i,\uparrow}  n_{i,\downarrow}  - \sum_{i,\sigma} \mu_{i} n_{i,\sigma}.
\label{HubbardHamiltonian}
\end{equation}
Here $c^\dagger_{i,\sigma} (c_{i,\sigma})$ denotes the fermionic creation (annihilation) operator at site $i$ with spin $\sigma$, $t$ is the nearest-neighbour tunnelling amplitude, $U < 0$ represents attractive on-site interaction and $\mu_i$ is the local chemical potential. We deploy the two lowest hyperfine states of $^{40}K$ in the $F=9/2$ ground state manifold, serving as the two spin states $\ket{\uparrow} = \ket{F=9/2, m_F = -9/2}$ and $\ket{\downarrow} = \ket{F=9/2, m_F = -7/2}$.

Two-dimensional planes are formed by an optical lattice with a lattice depth of $120\,E_r$, limiting tunnelling, where $E_r$ is the recoil energy of the optical light field. The in-plane square lattices are realized at a lattice depth of $6\,E_r$. Here $E_r = \frac{h^2}{8ma^2} = h \times \unit[4.41]{kHz}$ is the recoil energy, $h$ is the Plank's constant, $a= \unit[532]{nm}$ is the lattice spacing and $m$ is the atomic mass. The lattice depth is calibrated by performing lattice modulation spectroscopy, where the laser intensity of the individual lattice beam is periodically modulated. The resonant frequencies indicate the relevant band transitions and therefore offer calibration of the lattice depths. The combination of lattices give rise to an in-plane tunnelling amplitude  $t = h \times \unit[224(6)]{Hz}$. 

By tuning the magnetic field close to an s-wave Feshbach resonance near $202\,G$, we realize a wide range of attractive interactions $-1 > U/t > -10$ \cite{gall2019simulating}. 
We obtain the on-site interaction energy $U$ based on the analytical solution of two interacting fermions in an axially symmetric harmonic potential with an s-wave $\delta$-pseudopotential \cite{Idziaszek2006Analytical}. We then employ a lattice-depth dependent correction factor to account for the anharmonicity of the potential around a single lattice site \cite{Schneider2009} and obtain the values of $U/t$ presented in the manuscript. Using this calculation, we find a very nice agreement for the interaction energy shift between different hyperfine-state pairs measured in radio-frequency (RF) spectroscopy.

We note that we later utilize the local density approximation (LDA) in our analysis in which $\mu_i = \mu_0 - V(x,y)$, where $\mu_0$ is the chemical potential at trap center and $V(x,y)$ is the inhomogeneous confinement due to the optical lattice potential.  
To obtain this in-plane potential landscape $V(x,y)$, we measure the in-plane trap frequencies $\omega_x = 2\pi \times \unit[19.2(5)]{Hz} $ and $\omega_y = 2\pi \times \unit[25.9(3)]{Hz}$ by exciting the dipole oscillation in an non-interacting gas. From the measured trap frequencies, we can infer the relevant lattice beam waists $w_x = \unit[173(2)]{\mu m} $, $w_y =  \unit[152(1)]{\mu m} $ and $w_z =  \unit[105(1)]{\mu m}$. Using these parameters, together with the calibrated lattice depths, we can reconstruct the confinement $V(x,y)$ induced onto the atomic ensembles. 

%Temperatures of the ensembles are adjusted by varying the end point of evaporative cooling in the optical dipole trap. 

\begin{figure}[hb!]
\includegraphics[width=0.448\textwidth]{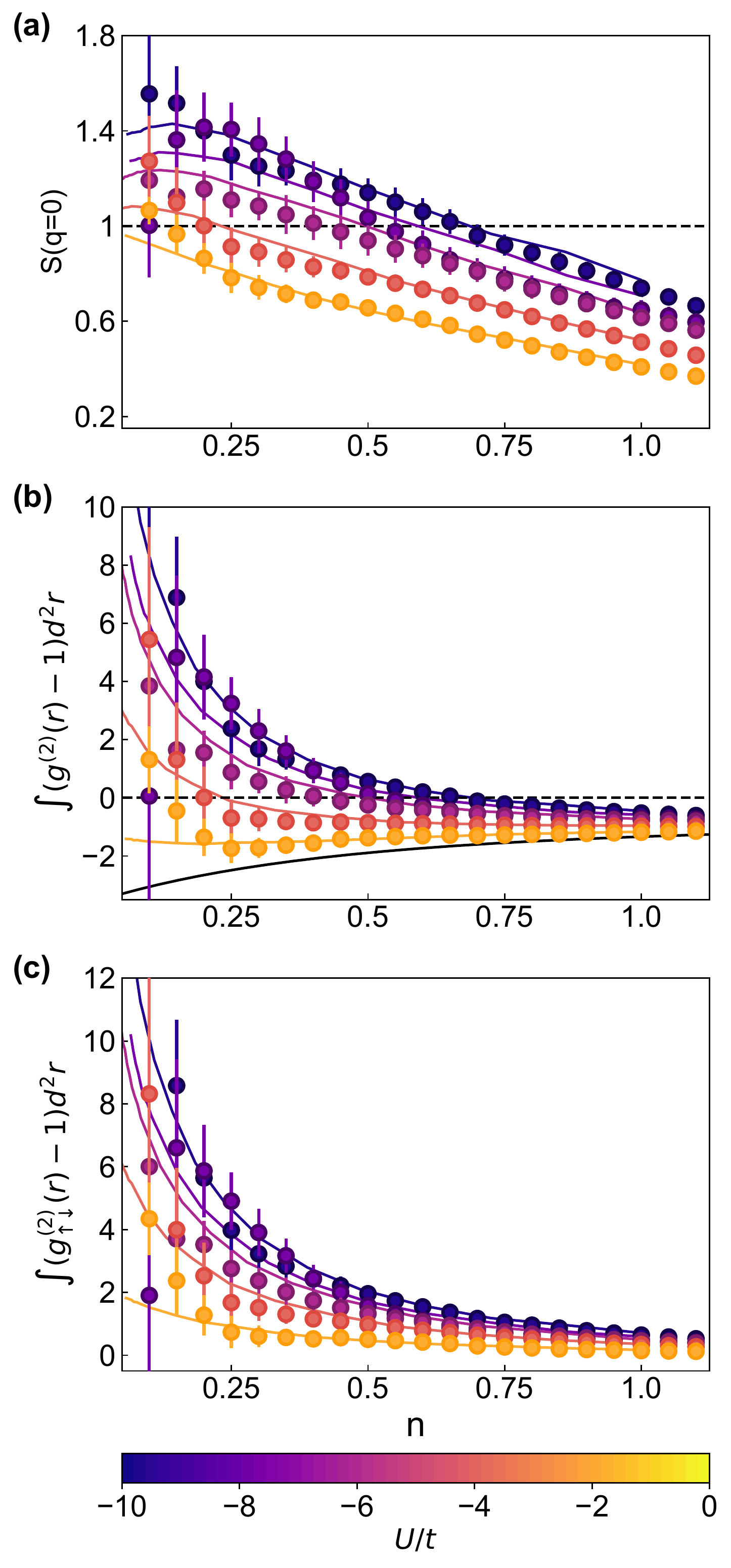}
\caption{(a) Density structure factor at zero momentum $S(\boldsymbol{q}=0)$ for various interaction strengths and filling. (b) $\int (g^{(2)} (\boldsymbol{r}) - 1) d^2r$ versus interaction strengths $U/t$. The black solid line shows the non-interacting expectation at the lowest temperature in the data sets. Data points above the horizontal dashed line signal that the unequal-spin $g_{\uparrow\downarrow}^{(2)}(\boldsymbol{r})$ outweighs its equal-spin counterpart, and vice versa. (c) Interacting contribution $\int (g_{\uparrow \downarrow}^{(2)} (\boldsymbol{r}) - 1) d^2r$ is obtained by subtracting the equal-spin contribution $\int (g_{\uparrow \uparrow}^{(2)} (\boldsymbol{r}) - 1) d^2r$. Data are for $U/t = -1.83$, $-3.90$, $-6.09$, $-7.62$ and $-9.61$, and the corresponding temperatures $k_BT/t$ of the data set are $1.26(4)$, $1.31(5)$, $1.73(8)$, $2.05(11)$ and $2.16(8)$ respectively. Solid lines show the result from DQMC simulation.}
\label{Fig2}
\end{figure}

\begin{figure*}[t]
\centering
\includegraphics[width=1.0\textwidth,clip=true]{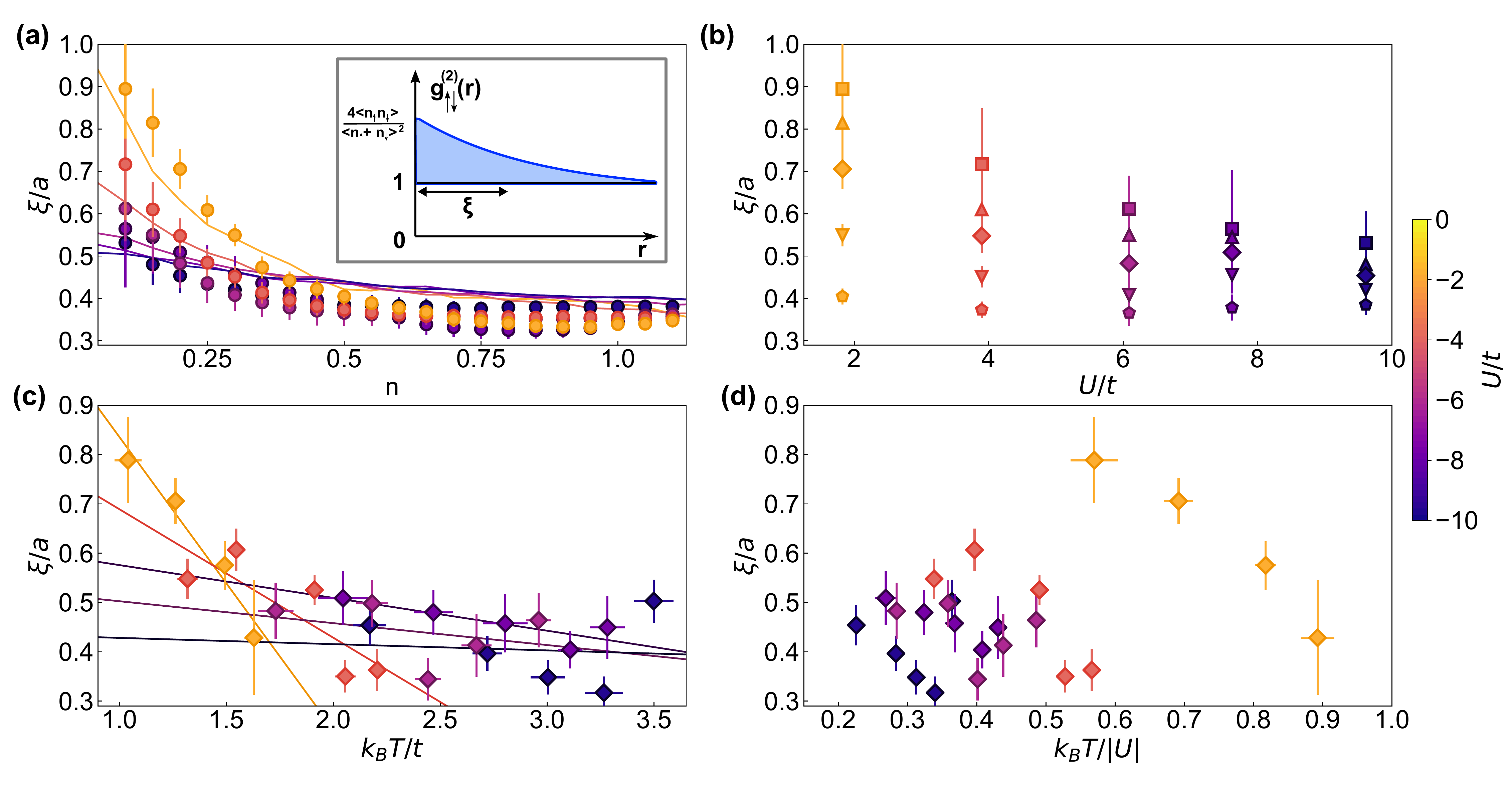}
\caption{(a)-(d) Inferred pair correlation length. (a) Correlation length $\xi$ versus filling $n$. As the interaction strength increases, we observe that the correlation length shrinks for low filling $n \lesssim 0.5$. For increasing filling, we observe that $\xi$ settles at approximately the limit of the continuous approximation. The inset exemplifies the estimation of the length scale, which is obtained by dividing area (integral) by height (amplitude). Solid lines show the result from DQMC simulation. (b) Correlation length $\xi$ for specific $n$ versus $U/t$. $n = 0.1$, $0.15$, $0.2$, $0.3$ and $0.5$ correspond to square, up-triangle, diamond, down-triangle and pentagon markers respectively. Temperatures of data points shown in (a) and (b) are same as those in Fig.~\ref{Fig2}. (c) Correlation length $\xi$ versus temperature $k_BT/t$ at $n = 0.2$. For low filling, we observe in general a decreasing trend as temperature rises. While already at $n = 0.5$, no significant temperature dependence is observed. The solid lines are linear fits to the data points. (d) Correlation length $\xi$ versus temperature $k_BT/|U|$ at $n=0.2$. When rescaled with respect to the interaction strength, the temperatures reached in the weakly-interacting case are higher than the ones in the strongly interacting case. }
\label{Fig3}
\end{figure*}

% Imaging 

The atomic ensembles are then locked in position by rapidly ramping up the in-plane lattice depths to $60\,E_r$ within~$\unit[1]{ms}$. A single layer of a two-dimensional sample is transferred to another internal state via radio-frequency (RF) tomography in the presence of a vertical magnetic field gradient. We then utilize the interaction shift between singly- and doubly-occupied states to further separate them into two hyperfine states for detection. Next, we deploy high-resolution absorption imaging to obtain the in-situ density profiles of singles $S_\uparrow = \langle n_\uparrow \rangle - \langle n_\uparrow n_\downarrow \rangle$ and doubles $D = \langle n_\uparrow n_\downarrow \rangle$. We exploit the knowledge of the optical potential $V(x,y)$ and use the local density approximation to determine the density equation of state $n(\mu)$ \cite{Cocchi2016} (also see Appendix A). Thermometry is performed via a {chi-squared} fitting of $n(\mu)$ with determinant quantum Monte-Carlo (DQMC) simulations \cite{Varney2009}, which allows us to extract the temperature $k_BT/t$  (see Appendix B).

Subsequently, we obtain the isothermal compressibility $\kappa = \partial n/\partial \mu$ through numerical differentiation of the measured equation of state $n(\mu)$. By taking the first derivative of a second-order polynomial fit to a subset of data-points in $n(\mu)$. The fit is performed on the data points over a chemical potential window of $h \times \unit[700]{Hz}$ around the desired $\mu$. For low filling, i.e. $n < 0.2$, we observe increasing technical noise due to the lower signal-to-noise ratio of the raw data. Thus we deploy an exponential fit in order to avoid fitting negative compressibility. In order to obtain the measurement at the same filling $n$, as presented in Fig.~2 and ~3 across different data sets, we interpolate neighbouring binned data points in $\mu$. 

In Fig.~\ref{Fig2}(a), we depict the density structure factor $S(\boldsymbol{q}=0)$ derived from the compressibility using the fluctuation-dissipation theorem. We observe that for low filling $S(\boldsymbol{q}=0)$ exceeds unity, which, quantitatively speaking, indicates particle-bunching. We note that $S(\boldsymbol{q}=0)$ in Eq.~(\ref{densitySF}) composes of both the equal-spin (Pauli) and unequal-spin (interacting) pair correlation functions. At fixed filling $n$, we observe that the structure factor increases monotonically with interaction strength, suggesting that the interacting pair correlation functions become increasingly dominant. 

We analyze the roles of both the equal- and the unequal-spin contributions by plotting the integral of the full $g^{(2)}(\boldsymbol{r})$ function, computed using Eq.~(\ref{g2Full}), as shown in Fig.~\ref{Fig2}(b). The horizontal dashed line indicates the point at which the equal and unequal-spin $g^{(2)}$ exactly compensate each other, \textit{i.e.} $\int \left[ g^{(2)} (\boldsymbol{r}) - 1 \right] d^2r = 0$ . The sign of this integral signals the dominant part in the pair correlation function. Thus, it offers a direct indication that the particle bunching observed in the density structure factor $S(\boldsymbol{q}=0)$ is caused by the fact that the attractive on-site interaction dominates over the Pauli blocking.

%Thus, it offers a direct indication that the particle bunching observed in the density structure factor $S(\boldsymbol{q}=0)$ is caused by the competition of Pauli blocking between particles with the same spin and the on-site attraction between those with opposite spins. 

%Similar to the bunching behavior of the structure factor, the sign flips occur at the same point as $S(\boldsymbol{q}=0)=1$. 
%While it is true that the tightly-bound local dimers would become composite bosons in the large-$U$ limit, our measurement scheme does not detect the pair-pair correlation. The bunching, as reflected in the $g^{(2)}(\boldsymbol{r})$, therefore, is not a sign for the bosonization of fermions.

In order to quantitatively compare the contributions from the Pauli principle and the attractive interaction, we note that the equal-spin contribution $\int \left[g_{\uparrow \uparrow}^{(2)} (\boldsymbol{r}) - 1\right] d^2r$ can be calculated using the tight-binding dispersion relation $\epsilon(k_x,k_y) = -2t \left[\cos(k_x a) + \cos(k_y a) \right]$ and temperature, see solid line in Fig.~\ref{Fig2}(b). Strictly speaking, this calculation is exact only for the non-interacting case because interactions in principle modify the dispersion from a simple sinusoidal energy band. However, in the low-filling regime, this estimate remains a faithful approximation since most of the occupied part of the energy band remains harmonic. Subtracting the equal-spin contribution from $\int \left[g^{(2)} (\boldsymbol{r}) - 1\right] d^2r$, we obtain the interacting $\int \left[g_{\uparrow \downarrow}^{(2)} (\boldsymbol{r}) - 1\right] d^2r$, as shown in Fig.~\ref{Fig2}(c). We observe that the interacting contribution maintains a similar dependence as $S(\boldsymbol{q}=0)$. For either decreasing filling or increasing interaction strength, the interacting pair correlation increases. This highlights the parameter space at which the interaction effect on $g^{2}_{\uparrow\downarrow} (\boldsymbol{r})$ is most prominent. At low filling, we observe a deviation of the measured pair correlations from the theoretical values, which we attribute to the low signal-to-noise near the low filling regime and the fact that the compressibility is vanishing for $n \rightarrow 0$ (vacuum). Both results in insensitivity of density with respect to trapping potential variation and thus leads to increased uncertainties in the compressibility.

To gain further insight into the signature of the pairing, we turn to estimate a length scale up to which the pair correlation extends. We start by noticing that the interacting pair correlation amplitude at $\boldsymbol{r} = 0$ is $g_{\uparrow\downarrow}^{(2)} (0) = 4 \frac{\langle n_{\uparrow}n_{\downarrow} \rangle}{\langle n_\uparrow+n_\downarrow\rangle^2} = \frac{D}{(S_\uparrow+D)^2}$. This implies that the amplitude can be directly obtained from our local density measurement of singly- and doubly-occupied site occupations. We note that in a lattice system, the double occupancy $D$ plays the role of the contact parameter, describing short-range pair correlation \cite{tan2008energetics,braaten2012universal}. Therefore, the non-trivial part of the pairing is reflected in the non-local part of the pair correlation function, which we analyze next.

 Although the analytical form of the unequal-spin pair correlation function is not known, an exponential decay $e^{-|\boldsymbol{r}|/\xi}$ is expected to be a good approximation above $T_c$, see inset of inset of Fig.~\ref{Fig3}(a). Combining the knowledge of the integral $\int \left[g_{\uparrow \downarrow}^{(2)} (\boldsymbol{r}) - 1\right] d^2r$ and the amplitude at $\boldsymbol{r}=0$, we then infer the characteristic length scale $\xi$ as
\begin{equation}
2\pi \xi^2  \approx  \frac{ \displaystyle{\int}f \left[g_{\uparrow \downarrow}^{(2)} (\boldsymbol{r}) - 1\right] d^2r}{\left[g_{\uparrow \downarrow}^{(2)} (0)  - 1\right]}.
\label{Estimate_xi}
\end{equation}

Eq.~(\ref{Estimate_xi}) renormalizes the measured pair correlation with respect to the on-site contribution. In Fig.~\ref{Fig3}(a), we plot the estimated pair correlation length $\xi$ as a function of filling $n$ for the same data set in Fig.~\ref{Fig2}. For low filling $n \lesssim 0.5$, we observe a correlation length $\xi$ as large as $0.92(4)\,a$ for the lowest interaction strength at $U/t=-1.83$, where $a = 532$\,nm denotes the in-plane lattice spacing. Although the attractive interaction, therefore the Hubbard model, is purely on-site, we observe that its effect extends beyond the local site, similar to the Pauli blocking leading to beyond-local density suppression. 

The trend of $\xi$ as a function of filling can be attributed to two reasons. First, for dilute filling, particles are described by delocalized wave-packets. As temperature decreases, particles tend to explore the bottom of the energy band. This scenario resembles the free-particle case. Therefore, in the superfluid phase, BCS pairs and BEC dimers are expected for weak and strong interaction strengths, respectively. We observe a qualitative agreement to this expectation, despite our temperature being higher than the critical temperature. With increasing filling, the band occupation and thus the influence of the interaction term increases. This results in reduced inter-particle spacing, and the latter drives the system away from the weak coupling limit despite the small $U$. Both contribute to the observed decrease in correlation length.

Second, the localization of particles at high filling means that the continuous integral of Eq.~(\ref{Estimate_xi}) starts to deviate from the discretized sum in a lattice. If the interacting contribution is dominated by the local term $g_{\uparrow\downarrow}(0)$, the continuous approximation of Eq.~(\ref{Estimate_xi}) would result in $\xi/a \approx \sqrt{1/2\pi} \approx 0.4$, which is in agreement with our observation in Fig.~\ref{Fig3}(a). Upon changing $U/t$, we observe that the correlation length $\xi$ shrinks as interaction strength increases, as shown in Fig.~\ref{Fig3}(b). This signals the formation of tightly-bound local pairs and this pairing behavior is most prominent below quarter filling ($n \lesssim 0.5$). Above quarter filling, we do not observe a discernible trend of $\xi$ as a function of $U/t$, as indicated by the lowest data points in Fig.~\ref{Fig3}(b).
 
 %In this case, the interacting contribution is dominated by the local term $g^{(2)}_{\uparrow\downarrow} (0)$, and Eq.~(\ref{Estimate_xi}) reduces to
 
Last but not least, we investigate the temperature dependence of the correlation lengths. For temperatures below the critical temperature and low density, the pairing would be described by a BCS-BEC type behavior. Although the temperatures reached in our experiments remain in the normal phase, we observe a resemblance in the behavior of the pair correlation lengths. As shown in Fig.~\ref{Fig3}(c), we plot $\xi$ as a function of temperature $k_BT/t$ at $n=0.2$. The correlation length rises as temperature decreases for the weakly attractive case. In the strongly attractive case, we observe a much less significant trend in temperature dependence. Since the pairing occurs at an energy scale of the interaction $U$, it is also informative to recast the temperature with respect to $|U|$. In Fig.~\ref{Fig3}(d), we compare the temperature with respect to the interaction energy by plotting the correlation lengths as a function of $k_BT/|U|$. Despite a lower achieved $k_BT/|U|$ for large interaction strengths, the correlation length remains small due to the energetically favorable dimer state. For weak interaction, the pair correlation length rises at much higher $k_BT/|U|$, signaling the tendency to delocalize and form longer-range pairs.

In conclusion, we investigate the formation of pair correlations across the crossover regime of a normal phase in the attractive Hubbard model. In particular, we observe the competition between Pauli repulsion and on-site attraction.  We show that for sufficiently low filling and weak interaction, the pair correlation length extends beyond local distance up to one site. This offers a clear signature for the formation of pairs as a potential precursor of the BCS-BEC crossover in a lattice configuration. Our measurement helps to elucidate the outstanding questions regarding the pairing behavior of a normal state. The approach presented here, in principle, also works beyond simple lattice geometries or on-site interaction, thereby allowing future investigation of long-range correlated systems. 
% Note: removed "A correlation length scale can be inferred from the data by subtracting the non-interacting contribution in the analysis." in the last paragraph!

This work has been supported by BCGS, the Alexander-von-Humboldt Stiftung, ERC (grant 616082), DFG (SFB/TR 185 project B4), Cluster of Excellence Matter and Light for Quantum Computing (ML4Q) EXC 2004/1 – 390534769 and Stiftung der deutschen Wirtschaft. 

\section{Appendix A: Thermometry}

\begin{figure}[h]
\includegraphics[width=0.5\textwidth]{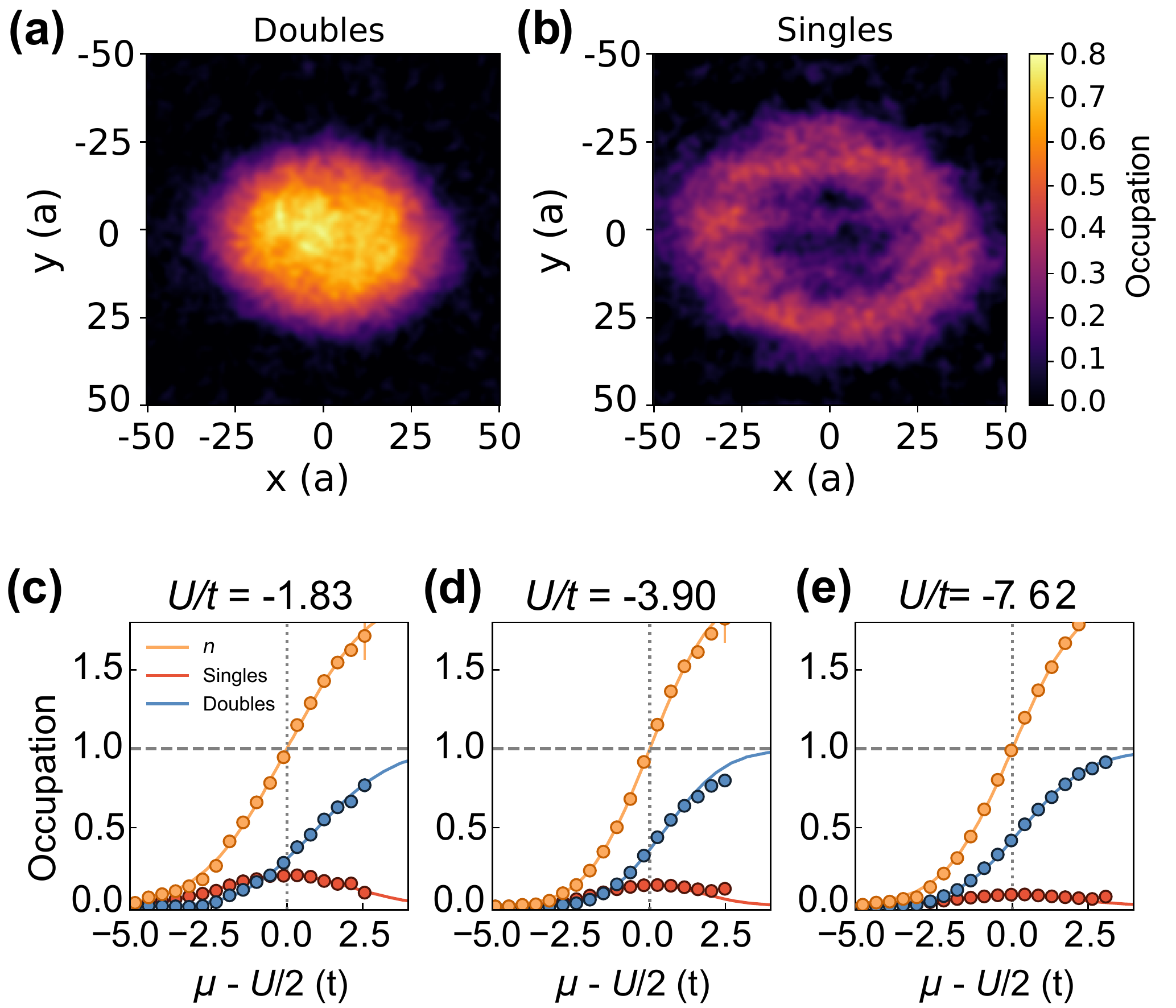}
\caption{(a) and (b) In-situ density profile of doubly-occupied states $D(x,y)$ and singly-occupied states $S_\uparrow(x,y)$, respectively. Images shown are obtained bt averaging over 40 experimental realizations with atom number fluctuations within $5\%$. (c)-(e) Density equation of state $n(\mu)$ for different interaction strengths. }
\label{sup-fig-1}
\end{figure}

Using the knowledge of $V(x,y)$, the recorded density profile $n(x,y)$ is mapped to the chemical potential axis under the local density approximation, \textit{i.e.} $\mu(x,y) = \mu_0 - V(x,y)$. This allows us to perform iso-potential averaging on in-situ profiles as shown in Fig. \ref{sup-fig-1} and to obtain both quantities as a function of local chemical potential $\mu$. The density equation of state $n(\mu)$ can then be computed using $n(\mu) = 2\left[S(\mu)+D(\mu) \right]$, where $S$ and $D$ are the occupation of singly-occupied state (\enquote{Singles}) and doubly-occupied states (\enquote{Doubles}), respectively. Here, we utilize the fact that we are spin-balanced, \textit{i.e.} $\langle n_\uparrow \rangle  = \langle n_\downarrow \rangle$. The global chemical potential $\mu_0$ and temperature $k_BT/t$ can be obtained by a numerical chi-square fit of $n(\mu)$ to the result of DQMC simulation.

\section{Appendix B: DQMC simulation}
For the theoretical prediction shown in the manuscript, we perform determinant quantum Monte Carlo (DQMC) simulation on an $8 \times 8$ 2D Hubbard model in a square lattice, using the Quantum Electron Simulation Toolbox (QUEST) Fortran package \cite{Varney2009}.  Simulations are performed with a wide range of interactions $-10 \leq U/t \leq -1$, with $1000$ warm-up sweeps and $50000$ measurement sweeps, and the number of imaginary time slices is set to $25$. Typically the average sign of the equal-time Green's function in DQMC exhibits a sign problem for lower filling, especially with strong interaction and low temperature. However, we numerically verify that the sign problem is negligible at our experimental temperature scales. The density structure factor $S(\bf{q}=0)$ can be obtained by a summation over all individual density-density correlators.

\end{document}